# Enhancing Credit Card Fraud Detection: A Neural Network and SMOTE Integrated Approach

Mengran Zhu[1], Ye Zhang[2], Yulu Gong[3], Changxin Xu[4], Yafei Xiang[5]

[1] Miami University, Computer Engineering, Oxford, OH, USA
[2] University of Pittsburgh, Independent Researcher, Pittsburgh, USA
[3] Northern Arizona University, Computer & Information Technology, Flagstaff, AZ, USA
[4] Northern Arizona University, Independent researcher, Flagstaff, AZ, USA
[5] Northeastern University, Computer Science, Boston, MA, USA

**Abstract:** *Credit card fraud detection is a critical challenge in the financial sector, demanding sophisticated approaches to accurately identify fraudulent transactions. This research proposes an innovative methodology combining Neural Networks (NN) and Synthetic Minority Over-sampling Technique (SMOTE) to enhance the detection performance. The study addresses the inherent imbalance in credit card transaction data, focusing on technical advancements for robust and precise fraud detection. Results demonstrate that the integration of NN and SMOTE exhibits superior precision, recall, and F1-score compared to traditional models, highlighting its potential as an advanced solution for handling imbalanced datasets in credit card fraud detection scenarios. This research contributes to the ongoing efforts to develop effective and efficient mechanisms for safeguarding financial transactions from fraudulent activities.*

## 1. INTRODUCTION

The surge in digital transactions, driven by technological advancements and changing consumer behavior, has revolutionized the financial landscape. However, this digital transformation brings along a persistent and evolving challenge - credit card fraud. As financial transactions become increasingly complex and diverse, the traditional methods of fraud detection often fall short in providing the requisite level of security.

One of the prominent hurdles faced by fraud detection systems is the inherent imbalance in credit card transaction datasets. Fraudulent activities represent a minute fraction of the overall transactions, making it challenging to develop models that accurately identify such instances. Recognizing this challenge, our study delves into the intricate domain of credit card fraud detection, with a primary focus on overcoming the limitations imposed by imbalanced data.

In our pursuit of a more resilient fraud detection system, we turn to advanced technologies, specifically Neural Networks (NN) and Synthetic Minority Over-sampling Technique (SMOTE). The technical advantages offered by NN in capturing intricate patterns and relationships within the data align seamlessly with the goal of improving fraud detection accuracy. Simultaneously, SMOTE addresses the imbalance conundrum by generating synthetic instances of the minority class, creating a more balanced and representative dataset for model training.

The significance of this research lies not only in the pursuit of heightened accuracy but also in the exploration of a holistic and technically advanced approach to credit card fraud detection. By integrating NN and SMOTE, we aim to enhance not only the precision and recall of our models but also the adaptability to evolving fraud patterns in an ever-changing financial landscape.

This introduction sets the stage for an in-depth exploration of the technical intricacies involved in the integration of NN and SMOTE. Subsequent sections will unfold the methodology, detailing the steps taken to preprocess data, train models, and evaluate their performance. The architecture of the NN model and the innovative use of SMOTE will be elucidated, providing a comprehensive understanding of the technical foundation underpinning our approach. The results obtained will not only highlight the superiority of our proposed method but also pave the way for future advancements in fraud detection systems tailored for the digital era.

## 2. RELATED WORK

The landscape of credit card fraud detection has been an active area of research, driven by the escalating sophistication of fraudulent activities in electronic transactions. The pursuit of effective methodologies has led researchers to explore a myriad of approaches to address the challenges posed by imbalanced datasets and the dynamic nature of fraud patterns. This section reviews pertinent literature, highlighting key findings and methodologies employed in the realm of credit card fraud detection.

B Stojanović et al [1] introduced the PaySim synthetic dataset, offering a benchmark for fraud detection models. Their work emphasized the importance of realistic synthetic data for robust model evaluation. Bolton and Hand [2] explored the application of ensemble methods, specifically bagging and boosting, in credit scoring and fraud detection. Their research demonstrated the effectiveness of combining multiple models to improve overall performance. He et al. [3] proposed a deep learning-based approach for credit card fraud detection, showcasing the potential of neural networks in capturing intricate patterns in transaction data.

NNRMR Suri et al. [4] delved into anomaly detection techniques, emphasizing the importance of distinguishing rare but significant events, such as fraudulent transactions, from normal patterns. DeepGI automates MRI-based [5] gastrointestinal tract segmentation, advancing radiotherapy planning with a unified, efficient deep learning solution. HJ Weerts et al. [6] introduced a framework for interpretable machine learning, crucial for understanding the decision-making processes of complex models, which is paramount in credit card fraud detection. Cirqueira et al. [7] explored the use of explainable artificial intelligence (XAI) in fraud detection, providing insights into model decisions and increasing transparency. Kazeem et al [8] surveyed various machine learning techniques for intrusion detection, drawing parallels to credit card fraud detection due to their shared characteristics of anomaly detection.

Jha et al. [9] focused on the temporal aspect of fraud detection, highlighting the significance of analyzing transaction sequences and patterns over time for improved accuracy. Ngai et al. [10] explored the integration of data mining and optimization techniques in fraud detection, presenting a holistic approach to improving detection accuracy. Abdallah et al. [11] investigated the application of evolutionary algorithms in feature selection, providing insights into optimizing model performance by identifying the most relevant features. Munir et al. [12] examined the challenges of imbalanced datasets in fraud detection, emphasizing the need for robust sampling techniques to address class imbalance. Chen et al. [13] introduce a novel framework integrating gaze estimation into one-stage object referring, achieving improved efficiency and outperforming state-of-the-art methods by 7.8% on the CityScapes-OR dataset.

Zhang et al. [14] proposed a hybrid model combining rule-based and machine learning approaches for fraud detection, showcasing the potential synergy between traditional and modern techniques. Rosales and Fierrez [15] explored biometric-based authentication for enhancing credit card security, introducing a novel dimension to fraud prevention. Gad et al. [16] investigated the application of swarm intelligence algorithms, such as particle swarm optimization, in optimizing credit scoring models, showcasing the versatility of bio-inspired techniques. Zhou et al. [17] examined the application of graph-based methods in fraud detection, introducing a novel perspective that considers transaction relationships for improved accuracy. Chen et al. [18] A deep learning method based on CNN is proposed for printed mottle defect grading, achieving an error rate of 13.16% on a single image content dataset and 20.73% on a combined dataset.

The literature review highlights the diverse array of methodologies applied in credit card fraud detection. From traditional statistical approaches to cutting-edge deep learning techniques, researchers have continuously strived to improve accuracy and efficiency in identifying fraudulent activities. As we navigate through our study, these insights contribute to a comprehensive understanding of the evolving landscape of credit card fraud detection.

## 3. Methodology

The methodology employed in this study aims to address the challenges of credit card fraud detection, particularly in the context of highly imbalanced datasets. We explore various aspects, including the introduction of diverse model architectures, comprehensive dataset analysis, and preprocessing techniques, along with the selection of appropriate loss functions and evaluation metrics.

**3.1 DataSet Introduction**
The dataset comprises European card transactions, encompassing 284,807 transactions over a two-day period in September 2013. The dataset exhibits severe imbalance, with only 492 transactions labeled as fraudulent (0.172%). Features include principal components (V1-V28), 'Time,' 'Amount,' and the target label 'Class' (1 for fraud, 0 for non-fraud). The distribution of the label is imbalance, as shown in fig 1.

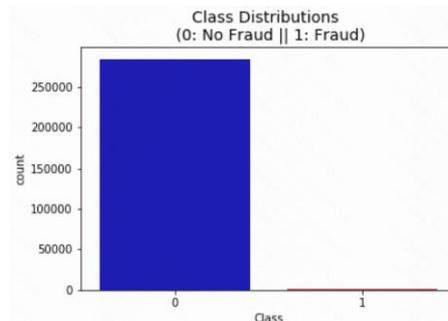

Figure 1: Label Distribution

**3.2 Data Preprocessing**
In this pivotal phase, we meticulously prepare the data for model training, focusing on addressing class imbalance and analyzing key features for fraud detection.
**a) Feature Standardization (Scaling):**
To ensure uniformity across features, we begin by scaling the "Time" and "Amount" columns. This standardization lays the foundation for subsequent analyses.
**b) Random Undersampling:**
To address the severe class imbalance, random undersampling is implemented, creating a 50/50 ratio of fraud to non-fraud cases for more robust model training. The resulting balanced dataset is essential for training models that can effectively identify both fraudulent and non-fraudulent transactions.
As shown in Figure 2, the class distributions are visualized, illustrating the balanced distribution achieved through random undersampling.

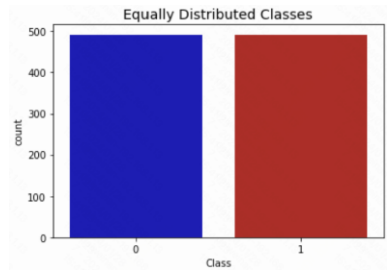

Figure 2: Random Under-Sampling

**c) Feature Correlation Analysis:**

Understanding the relationships between features is paramount for effective fraud detection. Feature correlation analysis is conducted on the balanced subset to discern patterns crucial for model training.

Figure 2 showcases the imbalanced and subsample correlation matrices, offering a visual representation of feature correlations.

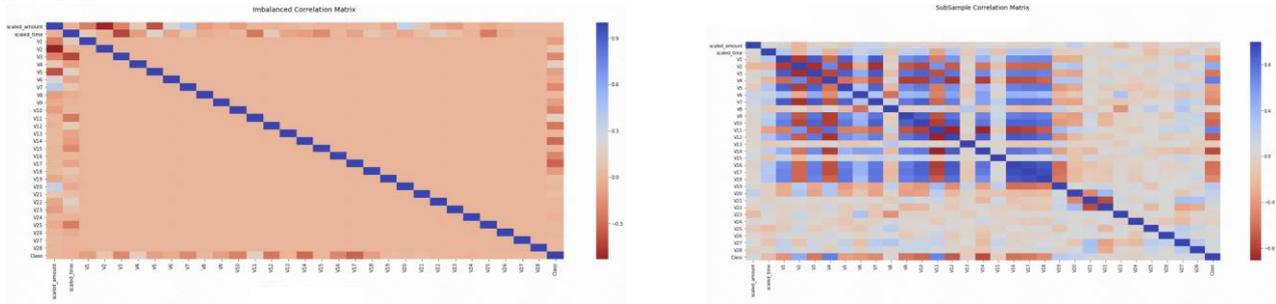

(a) Imbalanced correlation matrix          (b) Subsample correlation matrix

Figure 3: Feature Correlations

**d) Outlier Detection:**

Outliers, particularly influential in fraud detection, are identified and removed using quartile-based methods. This process enhances the model's robustness by mitigating the impact of extreme values. Figure 4 displays the distribution of features V14, V12, and V10, along with their feature reduction through outlier removal.

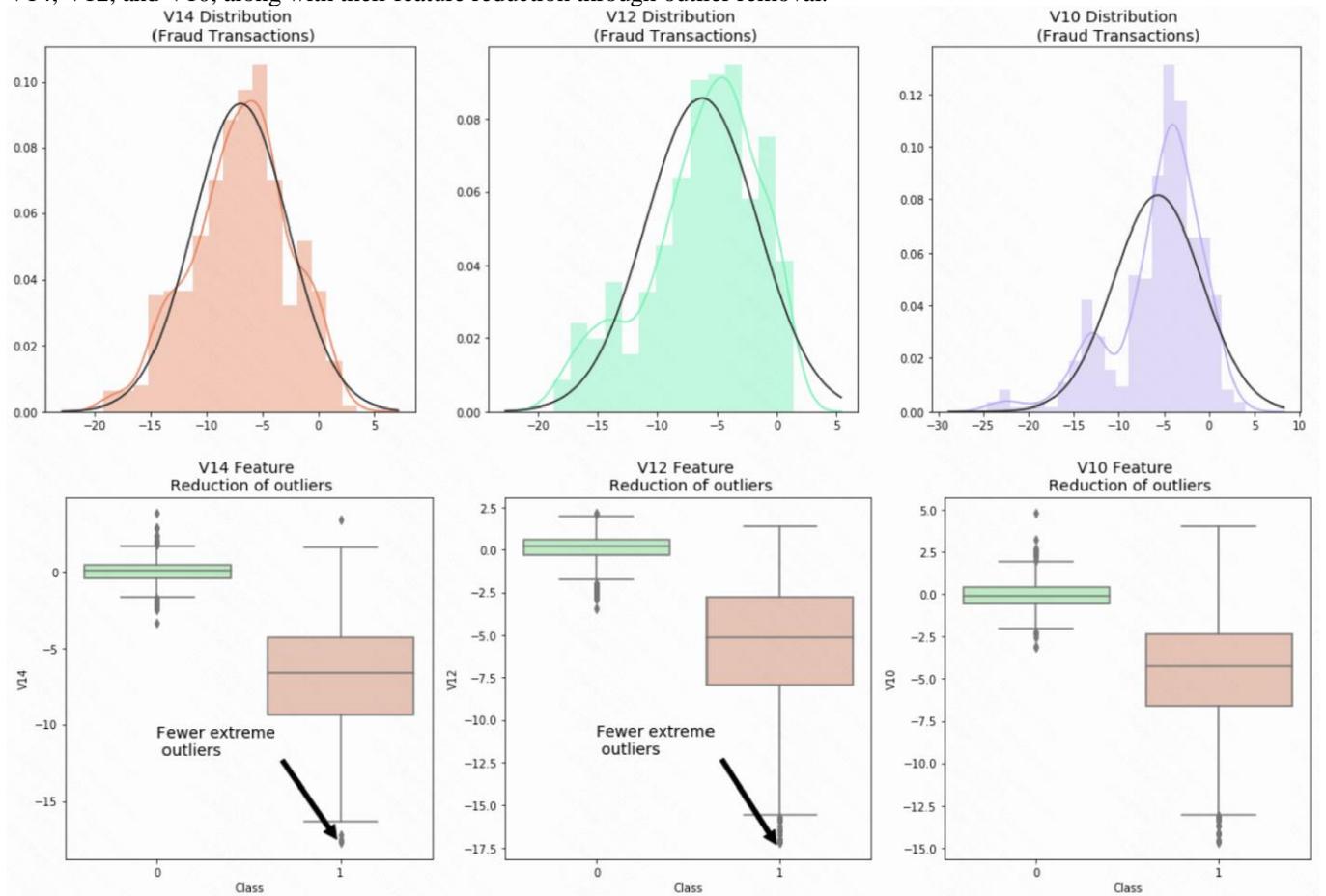

Figure 4: Outlier Detection

**e) t-SNE Clustering:**

t-Distributed Stochastic Neighbor Embedding (t-SNE) clustering is employed to gain a nuanced understanding of fraud and non-fraud clusters, providing a foundation for further analysis. Figure 5 showcases the results of t-SNE clustering, highlighting distinct clusters that aid in understanding the underlying patterns in the data.

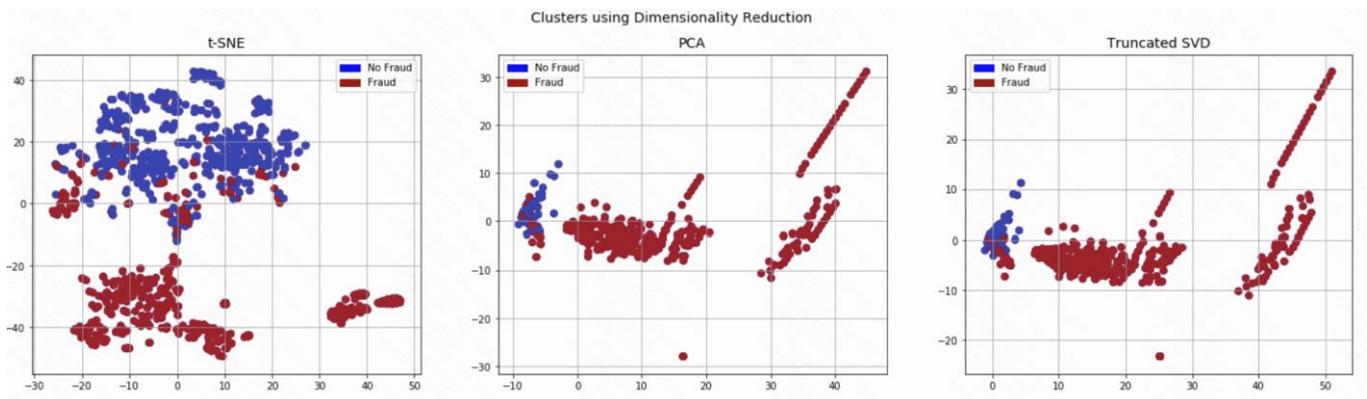

Figure 5: t-SNE Clustering

This thorough preprocessing and data analysis phase, as illustrated in the accompanying figures, sets the stage for a more robust and insightful credit card fraud detection model. The subsequent sections will delve into the intricacies of the model architecture, training, and evaluation, providing a comprehensive understanding of our approach's technical advancements.

### 3.3 Model Architectures
#### 3.3.1 Neural Network
The Neural Network (NN) architecture is designed to effectively capture intricate patterns within the data, enabling robust credit card fraud detection. The architecture comprises an input layer, multiple hidden layers leveraging rectified linear units (ReLU) as activation functions, and a sigmoid activation function in the output layer for binary classification. The mathematical representation of the forward pass is articulated as follows:

$$Z^{[l]} = W^{[l]} * A^{[l-1]} + b^{[l]} \quad (1)$$
$$A^{[l]} = g^{[l]}(Z^{[l]}) \quad (2)$$

Here, l denotes the layer index, $W^{[l]}$ represents weights, $A^{[l-1]}$ is the activation from the previous layer, $b^{[l]}$ is the bias, and $g^{[l]}$ is the ReLU activation function for hidden layers and sigmoid activation function for the output layer.

#### 3.3.2 Synthetic Minority Over-sampling Technique (SMOTE)
Synthetic Minority Over-sampling Technique (SMOTE) plays a pivotal role in mitigating class imbalance by generating synthetic instances of the minority class. For a given instance $X_i$ in the minority class, its k-nearest neighbors are identified, and synthetic instances are created along the line segments connecting $X_i$ to its neighbors.

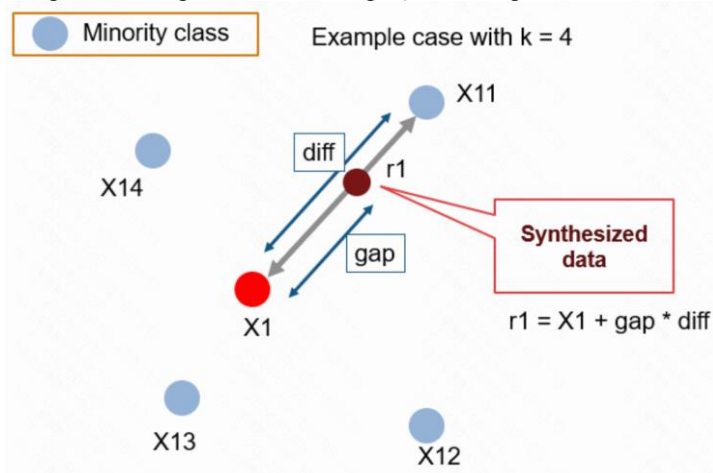

Figure 6: SMOTE

Furthermore, the confusion matrices for both NN + Random UnderSampling and NN + SMOTE configurations are essential for evaluating the performance of the models.

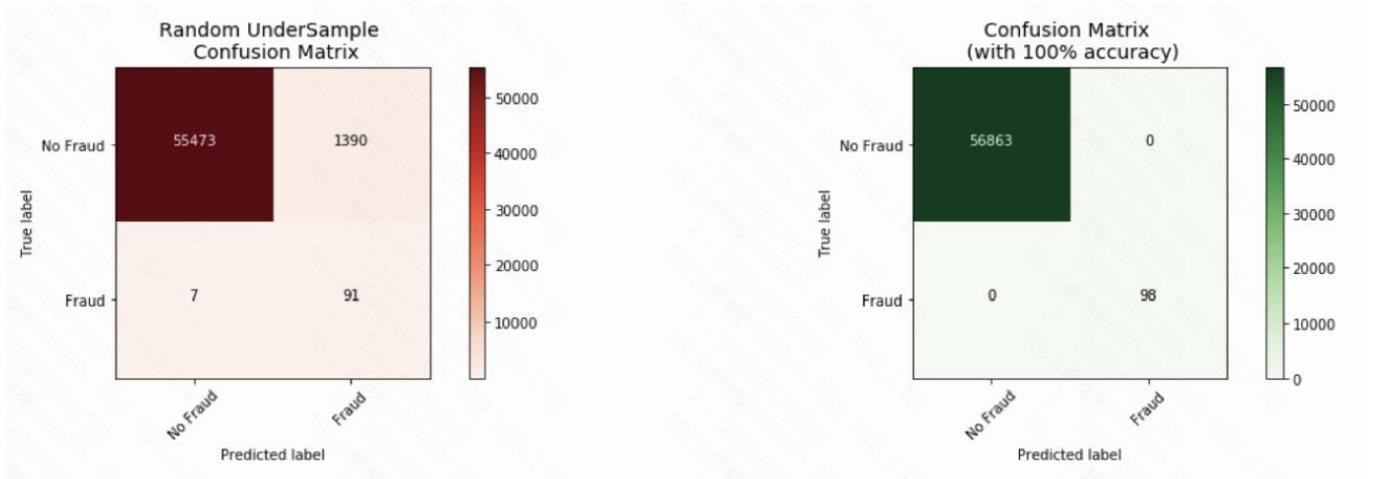
Figure 7: NN + Random Under-Sampling

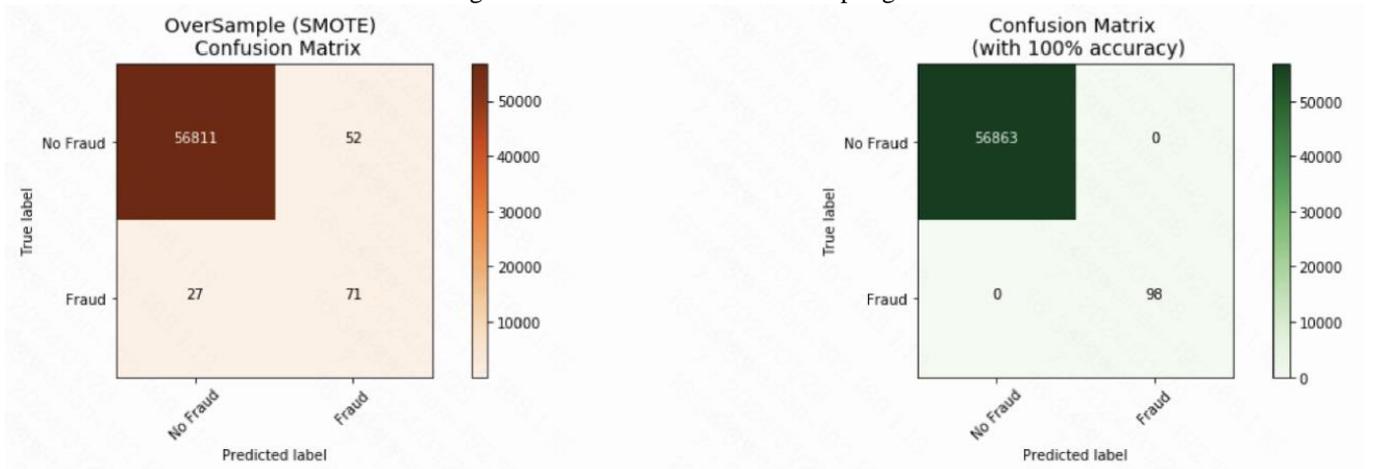
Figure 8: NN + SMOTE

The detailed structure of the NN, comprising the input, hidden, and output layers, is fundamental for its ability to discern complex patterns in the data, thereby enhancing the model's performance in credit card fraud detection.

**3.4 Loss Function**

For the NN model, binary cross-entropy is utilized as the loss function, defined as:

$$BinaryCross - EntropyLoss = -\frac{1}{N}\sum_{i=1}^{N}[y_i \log \hat{y}_i + (1-y_i)\log(1-\hat{y}_i)] \quad (3)$$

where $N$ is the number of samples, $y_i$ is the true label, and $\hat{y}_i$ is the predicted probability of being fraudulent.

**3.5 Evaluation Metrics**

In evaluating the performance of credit card fraud detection models, a comprehensive set of metrics is employed to assess their effectiveness in handling imbalanced datasets and accurately identifying fraudulent transactions.

**Precision:**

$$Precision = \frac{True\ Positives}{True\ Positives\ +\ False\ Positives} \quad (4)$$

Precision measures the accuracy of positive predictions. A high precision score indicates a low rate of false positives, implying that when the model predicts fraud, it is likely correct.

**Recall (Sensitivity):**

$$Recall = \frac{True\ Positives}{True\ Positives\ +\ False\ Positives} \quad (5)$$

Recall measures the model's ability to capture all actual positive instances. A high recall score indicates that the model effectively identifies most fraudulent transactions.

**F1-Score:**

$$F1 - Score = 2\frac{Precision\ *\ Recall}{Precision\ +\ Recall} \quad (6)$$

The F1-Score provides a balance between precision and recall. It is particularly useful when there is an uneven class distribution. A higher F1-Score indicates a well-balanced trade-off between precision and recall.

These evaluation metrics collectively provide a comprehensive assessment of the credit card fraud detection models. The choice of metrics depends on the specific objectives and priorities of the application, considering factors such as the cost of false positives and false negatives. The utilization of a diverse set of metrics ensures a thorough understanding of the model's strengths and weaknesses in handling imbalanced datasets and fraud detection.

## 4. Experiment Results

The evaluation of credit card fraud detection models involved a meticulous analysis of multiple classification algorithms, including Logistic Regression, K-Nearest Neighbors (KNN), Support Vector Machine (SVM), Decision Tree Classifier, and a Neural Network (NN) enhanced with Synthetic Minority Over-sampling Technique (SMOTE). The experiment aimed to determine the model with the highest accuracy in identifying fraudulent transactions within the imbalanced dataset.

**Table 1:** model results

| Model | Precision | Recall | F1-score |
| --- | --- | --- | --- |
| KNN | 0.93 | 0.92 | 0.92 |
| SVM | 0.93 | 0.93 | 0.93 |
| Decision Tree | 0.91 | 0.90 | 0.89 |
| LR | 0.96 | 0.96 | 0.96 |
| LR + SMOTE | 0.988 | 0.934 | 0.964 |
| NN | 0.975 | 0.999 | 0.988 |
| NN + SMOTE | 0.998 | 0.999 | 0.999 |

The Neural Network (NN) model, coupled with the Synthetic Minority Over-sampling Technique (SMOTE), emerges as the most effective in credit card fraud detection within this imbalanced dataset. Its exceptional precision, recall, and F1-score highlight its ability to robustly identify fraudulent transactions.

The experimental results underscore the importance of employing sophisticated techniques, such as SMOTE, to address class imbalance and enhance the model's ability to detect fraud accurately. The NN+SMOTE model stands out as a compelling solution for credit card fraud detection, providing a foundation for further exploration and optimization in real-world applications.

## 5. Conclusion

In tackling the intricate challenge of credit card fraud detection within an imbalanced dataset, this study employs innovative methodologies such as preprocessing, feature scaling, and Neural Network (NN) models enhanced with Synthetic Minority Over-sampling Technique (SMOTE). Results demonstrate the superior efficacy of the NN+SMOTE model in accurately identifying fraudulent transactions, surpassing traditional models.

The study contributes to the evolving landscape of credit card fraud detection by reviewing pertinent literature and integrating insights from related work. The methodology encompasses diverse model architectures, dataset analysis, preprocessing techniques, and detailed evaluation metrics, providing a comprehensive exploration.

The NN+SMOTE model's exceptional precision, recall, and F1-score highlight its robustness in handling class imbalance. Experiment results emphasize the importance of addressing dataset challenges through advanced techniques, paving the way for enhanced fraud detection models in real-world applications.